\documentclass[12pt,a4paper]{article}

\usepackage{amsmath}
\usepackage{amsfonts}
\usepackage{amssymb}
\usepackage{graphicx}
\usepackage{color}

\newcommand{\be}{\begin{eqnarray}}
\newcommand{\ee}{\end{eqnarray}}

\newcommand{\sign}{\mathrm{sign}}
\newcommand{\rp}{\mathrm{p}}

\newcommand{\bp}{\mathbf{p}}
\newcommand{\bq}{\mathbf{q}}

\newcommand{\ZZ}{\mathbb{Z}}

\renewcommand{\bar}{\overline}
\renewcommand{\hat}{\widehat}

\newcommand{\p}{\partial}
\newcommand{\half}{\frac{1}{2}}

\renewcommand{\(}{\left(}
\renewcommand{\)}{\right)}

\renewcommand{\l}{\langle}
\renewcommand{\r}{\rangle}

\begin{document}
\begin{center}
{\Large \textbf{Models for $d$ Wave Topological Superconductors and Quantum Anomalous Hall Effect with Arbitrary Large Chern Numbers}}\\
\bigskip
\smallskip
\centerline{Tong Chern$^\dagger$}
\smallskip
\centerline{\it School of Science, East China University of Technology, Nanchang 330013, China}
\medskip \centerline{\it $^\dagger$ 1643399509@qq.com}
\end{center}

\begin{abstract}

We construct theoretical models for two dimensional(2d) chiral $d_{x^2-y^2}\pm id_{xy}$ topological
superconductors and for three dimensional(3d) $d$ wave topological superconductors.
Moreover we build models for any 2d class C and 3d class CI topological superconductors
(with any even topological invariants).
We also construct concrete models that can realize quantum anomalous Hall effect
with arbitrary large Chern numbers. We study the chiral edge modes
or gapless surface states of our 2d or 3d models in details.
In all the cases, we find novel mechanisms that make the numbers of boundary states
always agree with the nontrivial bulk topology,
just as required by the bulk boundary correspondence.


\end{abstract}

\tableofcontents

\bigskip
\bigskip
\bigskip

\section{Introduction}

~~~~~In recent few years, there has been a surge of interest
in the study of the topological phases of free fermions\cite{TI1}\cite{TI2}\cite{TI3}.
In these phases, the system is fully gapped in bulk and has a unique ground state
on a compact $d$ dimensional spatial
manifold $M$ without boundary. If $M$ has a boundary,
the system will have gapless excitations along this boundary.
And there is a topological classification for these free fermion systems\cite{kane1}\cite{kane2}\cite{kane3}\cite{3d1}\cite{3d2}.
In this classification, one groups the Hamiltonians that can be smoothly deformed into each others without closing
the bulk gap as a homotopical class. The systematical classifications show a regular pattern as a
function of symmetry class and dimensionality, and can be arranged into a periodic table\cite{pt1}\cite{pt21}\cite{pt2}\cite{pt3}.

 As for the topological superconductors (in two dimensions(2d) and three dimensions(3d)) of the periodic table,
 many have been known for the cases with both conventional time reversal(TR) symmetry (with $T^2=-1$)
 and conventional particle hole symmetry (with $C^2=1$), see \cite{TI1}\cite{TI2} and references there in.
 Also the TR breaking case in 2d with conventional particle hole symmetry ($C^2=1$)
 has been extensively researched \cite{read}.
 But, so far, it has been largely overlooked
 for the 2d TR breaking $C^2=-1$ (class $\mathrm{C}$ in the Altland-Zirnbauer(AZ) classification \cite{az1}\cite{az2})
 cases and for the 3d TR invariant cases with $T^2=1$ and $C^2=-1$ (class $\mathrm{CI}$).

On the other hand, chiral superconductivity with a $d_{x^2-y^2}\pm id_{xy}\ (d+id)$ orbital symmetry
is proposed by \cite{dwave1}\cite{dwave11}\cite{dwave2}\cite{dwave20}\cite{dwave21}\cite{dwave3}.
And, this degeneration of the $d_{xy}$ and $d_{x^2-y^2}$ ordered states has recently
been demonstrated by Nandkishore et al.\cite{dwave4} in doped graphene monolayer.
It was shown \cite{dwave1}\cite{dwave11}\cite{dwave2}\cite{dwave3} that,
in these $d_{x^2-y^2}\pm id_{xy}$ superconductors, there are gapless chiral edge modes and
a quantized spin (should not be confused with the quantum spin Hall effects)
and thermal Hall conductance, which signal the nontrivial topology of the $(d+id)$ state.
Thus, certain 2d $d+id$ superconductors may be classified as the Class C topological superconductors.

 One of the purposes of the present paper is to construct nontrivial models that can realize
 the class $\mathrm{C}$ and class $\mathrm{CI}$ topological superconductors with any possible topological invariants.
 For 2d class $\mathrm{C}$, our constructions, in the simplest case,
 turn out to be the models with $d_{x^2-y^2}\pm id_{xy}\ (d+id)$ pairing symmetry.
We compute the topological invariants of these models, and derive a TKNN formula that relate the Chern numbers
to the spin quantum Hall conductance (should not be confused with the quantum spin Hall effects).
The dissipationless chiral edge excitations of our models have been studied,
with the results in agreement with \cite{dwave1}\cite{dwave11}.
Besides, we also construct concrete $d$ wave superconductor models for
the 3d class $\mathrm{CI}$ topological superconductors, and study the surface states of
these models in details.

Another purpose of the present paper is to explicitly construct
models for quantum anomalous Hall effect with arbitrary large Chern numbers.

The quantum anomalous Hall effect (QAH)\cite{QAr} is the quantum Hall effect in some ferromagnetic insulators,
which can have quantized Hall conductance (characterized
by first Chern number $c_1$), contributed
by dissipationless chiral states at sample edges, even without external magnetic field\cite{haldane}.
Recently, the QAH effect has been experimentally discovered in magnetic topological insulator of $\mathrm{Cr}$-doped
$\mathrm{(Bi,Sb)_2Te_3}$, where the $c_1=1$ has been reached\cite{qa}. Search for QAH insulators with higher Chern numbers
could be important both for fundamental and practical interests, since the QAH effect with higher
Chern numbers can lower the contact resistance, and significantly improve the performance of the interconnect devices.
In recent years, there are many interests in this direction\cite{hc0}\cite{hc1}\cite{hc2}\cite{hc3}\cite{hc4}.
In the present paper, we will explicitly construct QAH models that can realize arbitrary large Chern numbers.
It turns out that in the $c_1=1$ case,
our models become the model introduced in \cite{qam}(under taking the continuum limit).
We studied the chiral edge modes of our generalized QAH models,
which have some novelties that does not appear in the $c_1=1$ case.

As it is well known, the QAH states are deeply related to the two dimensional quantum spin Hall effect\cite{qa2d}
and to the 3d TR invariant topological insulators\cite{qa3d}, see\cite{qare}for a review. We can
thus construct a class of new models
for quantum spin Hall effects and for 3d TR invariant topological insulators. These new models are closely related
to our generalized QAH models. In the simplest cases, our models for quantum spin Hall effect are identical to
the Bernevig, Hughes and Zhang (BHZ) model\cite{BHZ}. Our models for 3d TR invariant topological insulators can also be seen as a
natural generalization of the low energy effective model for $\mathrm{Bi_2Se_3}, \mathrm{Bi_2Te_3}$ and $\mathrm{Sb_2Te_3}$, introduced by\cite{TI}.
But the boundary states of our new models exhibit some novel phenomena that can not be seen in the simplest cases.

The present paper is organized as follows. In section 2,
we construct simple theoretical models for 2d chiral $d+id$ topological superconductors (class C).
The topological invariants of these models are carried out by utilizing a homotopical argument.
In this section we also construct models for three dimensional $d$ wave topological superconductors (class CI). In section 3,
we present our constructions for QAH models with arbitrary Chern numbers. In section 4,
a large class of new models for $Z_2$ TR invariant topological insulators (in 2d and 3d) are constructed.
Section 5 study the chiral edge modes of our models for QAH effects with large Chern numbers
and study the chiral edge models of the likewise models for 2d TR breaking $d+id$ topological superconductors.
Section 6 study the subtle problem of surface states of our new models for 3d strong topological insulators.
Section 7 study the surface states of our models for three dimensional class $\mathrm{CI}$ topological superconductors.
In the last section, we discuss some possible applications of our theory.

\section{Simple Models for $d$ Wave Topological Superconductors}

~~~~In this section, we will construct a simple model for 2d chiral $d+id$ superconductor,
our model is different with the model introduced in \cite{vivo}
and can be easily generalized to the whole symmetry class (class C).
We can demonstrate the existence of a topologically nontrivial phase in our model, by explicitly
calculating the Chern numbers. Further more, in the present section,
we will construct a topologically nontrivial model for 3d $d$ wave superconductor also.

Our $d+id$ superconductor model is described by the following mean field theory Hamiltonian
\be H=\half\sum_{{\bp}}\Psi^{\dagger}_{\mathrm{\bp}}H_{BdG}(\bp)\Psi_{\mathrm{\bp}}.\label{ddmodel}\ee
Where $\Psi_{\mathrm{\bp}}=(c_{\mathrm{\bp}\uparrow}, c^{\dagger}_{-\mathrm{\bp}\downarrow})^{T}$ is the Nambu spinor,
and the Bogoliubov de Gennes (BdG) Hamiltonian $H_{BdG}$ for the quasiparticles is given by
\be
H_{BdG}(\bp)=\(\begin{array}{cc}
t(p^2_x+p^2_y)^2-\mu & \bar{\Delta}(p_x-ip_y)^2\\
\Delta(p_x+ip_y)^2 & -(t(p^2_x+p^2_y)^2-\mu)
\end{array}\)\label{bdg}\ee
where $t>0$, and $\Delta(p_x+ip_y)^2$ is the $d+id$ pairing function.
This model is apparently TR broken,
but preserves particle hole symmetry $C$, which acts on the Nambu spinor as
\be C=\(\begin{array}{cc}
0 & -1\\
1 & 0
\end{array}\)K,\ee
where $K$ stands for complex conjugation. Obviously, $C^2=-1$.
Hence our model is in type of class C in terms of AZ classifications.

The diagonal quartic terms of (\ref{bdg}) may make our model
somewhat peculiar, but they are necessary (to give a nontrivial
Chern number $c_1=2$, as we will see).
In a lattice with unit lattice constant, the quartic term $t(p^2_x+p^2_y)^2$ may be realized
as $4t(2-\cos(p_x)-\cos(p_y))^2$, and the pairing function $\Delta(p_x+ip_y)^2=\Delta(p_x^2-p_y^2+2ip_xp_y)$ may be realized as
\be 2\Delta\((\cos(p_x)-\cos(p_y))+i\sin(p_x)\sin(p_y)\),\ee
 which is a complex mixing of
$d_{x^2-y^2}$ wave and $d_{xy}$ wave.

The two eigenvalues of $H_{BdG}(\bp)$ are
\be E(\bp)_{\pm}=\pm\sqrt{(t\bp^4-\mu)^2+|\Delta|^2\bp^4}.\ee
Obviously, if $\mu\neq 0$, our system is fully gapped at the whole momentum space.
And we notice that when $\bp\rightarrow 0$, $E(\bp)_{\pm}\rightarrow \pm\mu$. Hence when we continuously vary $\mu$, from negative to positive,
the two energy "bands" are inverted. This "bands" inversion indicate a topological phase transition.
In fact, for $\mu<0$, $H_{BdG}(\bp)$ is topologically trivial, since in this case we can continuously deform
the parameter $\Delta$ to 0 (the $\Delta=0$ system is of course trivial), without closing the energy gap. While for the case of $\mu>0$, the system is indeed topologically nontrivial. In what follows we will prove this claim by combing a homotopical argument with an explicit calculation to the Chern number $c_1$.

Now we continuously deform the various parameters of $H_{BdG}(\bp)$, but holding $\mu>0$, $t>0$ and $|\Delta|>0$.
There is no topological phase transitions in these deformations, since the energy gap is always maintained.
All the Hamiltonians that can be deformed into each other are homotopically equivalent,
and will have the same topological invariants.
Hence, if $\mu>0$, $H_{BdG}(\bp)$ is homotopically equivalent to
\be
\(\begin{array}{cc}
|\Delta|^2(p^2_x+p^2_y)^2-1 & 2\bar{\Delta}(p_x-ip_y)^2\\
2\Delta(p_x+ip_y)^2 & -(|\Delta|^2(p^2_x+p^2_y)^2-1)
\end{array}\).\label{bdgt}\ee

One can easily find the eigenstates of (\ref{bdgt}). We will denote the eigenstate with eigenvalue $E_-(\bp)=-(1+|\Delta|^2\bp^4)$
as $\Phi(\bp)$,
\be\Phi(\bp)=\(\begin{array}{cc}
 1\\
-\Delta(p_x+ip_y)^2
\end{array}\)/\sqrt{1+|\Delta|^2\bp^4}.\ee
One can then calculate the Berry phase $\mathcal{A}=i\Phi^{\dagger}(\bp)d\Phi(\bp)$ and Berry curvature $\mathcal{F}=d\mathcal{A}$,
with, $\mathcal{F}=8\frac{|\Delta|^2\bp^2}{(1+|\Delta|^2\bp^4)^2}$.
Thus the Chern number $c_1$ can be calculated
\be c_1=\int_{\bp}\frac{\mathcal{F}}{2\pi}=\frac{4}{\pi}\int_{\bp}\frac{|\Delta|^2\bp^2}{(1+|\Delta|^2\bp^4)^2}dp_xdp_y=2.\label{c1}\ee
Hence, when $\mu>0$, our $d+id$ superconductor model (\ref{bdg}) describe a topological superconductor with $c_1=2$.
On the other hand, since our model (\ref{bdg}) is in class C symmetry type, it should be a 2d class $\mathrm{C}$ topological superconductor.
And it is well known that this class of topological superconductors is classified by $2\ZZ$ rather than by $\ZZ$\cite{pt3}.
This is in agreement with our explicit calculation to $c_1$ (\ref{c1}).

The topological invariant $c_1$ can be related to a physical observable,
the Hall spin conductance,
which is an analog of the Hall conductance. To give a precise definition,
we notice that the Nambu spinor $\Psi_{\bp}$ has definite spin $\hbar/2$, and thus our model (\ref{ddmodel})
is invariant under the spin rotation around $z$ axis,
\be \Psi_{\bp}\rightarrow e^{i\frac{\hbar}{2}\theta}\Psi_{\bp}.\ee
This is a global $U(1)$ symmetry with 'charge' $e=\hbar/2$,
thus we have conserved spin current
\be J_{\bq}=\sum_{\bp}\Psi^{\dagger}_{\bp+\bq/2}\frac{\p H_{BdG}}{\p \bp}\Psi_{\bp-\bq/2},\ee
and spin density
\be \rho_{\bq}=\sum_{\bp}\Psi^{\dagger}_{\bp+\bq}\Psi_{\bp}.\ee
This spin density is coupled to the Zeeman field $B_z(x)$. Thus
one can define the Hall spin conductance\cite{dwave11} $\sigma^{s}_{xy}$ by,
\be J_y=\sigma^{s}_{xy}(-dB_z(x)/dx).\ee
The well known TKNN formula\cite{TKNN} will then tell us
\be \sigma^{s}_{xy}&=&-\frac{e^2}{h}c_1=-\frac{(\hbar/2)^2}{h}c_1\\
&=&-\frac{\hbar}{8\pi}c_1=-2\cdot\frac{\hbar}{8\pi}.\ee
This quantization of the Hall spin conductance is firstly derived in \cite{dwave11},
by using a different method.

Having provide a nontrivial model for 2d class $\mathrm{C}$ topological superconductor with $d+id$ paring symmetry,
it is natural to try to further construct a $d$ wave model for 3d class $\mathrm{CI}$ topological superconductor.
Besides the particle hole symmetry $C^2=-1$, this model should be TR invariant (with $T^2=1$) at the same time.
We can explicitly give this model,
in the four components Nambu spinor basis $\Psi(p)=\(c_{\rp\uparrow},c^{\dagger}_{-\rp\downarrow},ic_{\rp\downarrow},ic^{\dagger}_{-\rp\uparrow}\)^{T}$, as
\be
H_{BdG}=\(\begin{array}{cccc}
E(p) & \bar{\Delta}(p_x-ip_y)^2 &0& -\Delta_z p_z\\
\Delta(p_x+ip_y)^2 & -E(p) & \Delta_z p_z&0\\
0& \Delta_z p_z & E(p) & \Delta(p_x+ip_y)^2\\
-\Delta_z p_z & 0& \bar{\Delta}(p_x-ip_y)^2 &-E(p)
\end{array}\),\label{3dhamil}\ee
where, $E(p)=t(p^2_x+p^2_y)^2+t_zp^{2}_z-\mu $, with $t>0, t_z>0$, and $\Delta_z$ is real.
At first sight this model seems to be TR breaking, since the pairing function is a complex function.
But this is in fact due to the particular choice of basis, in deed, there is a TR symmetry, it acts as (in the present basis)
\be T=-i\(\begin{array}{cccc}0&0&1&0\\
0&0&0&1\\
1&0&0&0\\
0&1&0&0
\end{array}\)K=-i\sigma_xK,\ee
where $\sigma_x$ is the Pauli matrix acts on the spinor indices.
And the particle hole symmetry acts as
\be C=\(\begin{array}{cccc}0&-1&0&0\\
1&0&0&0\\
0&0&0&-1\\
0&0&1&0
\end{array}\)K=-i\tau_yK,\ee
where $\tau_y$ is the Pauli matrix acts on the two components Nambu pseudo spinor indices.

Just as in the 2d case,
one can show that (\ref{3dhamil}) is gapped at the whole momentum space if $\mu\neq 0$. If $\mu<0$, one can continuously deform the parameters
$\Delta$ and $\Delta_z$ to zero, without closing the gap, thus the model will be in the topologically trivial phase. At $\mu=0$, there is a quantum phase transition. When $\mu>0$, the system will transit to a topologically nontrivial phase. In section (7), we will show (by studying the gapless surface states) that there is a nontrivial topological invariant $2$ characterising this $\mu>0$ phase. This is in agreement with the $2\ZZ$ classification for 3d class $\mathrm{CI}$ topological superconductors \cite{pt3}.

We would like to note that our model for class $\mathrm{CI}$ topological superconductor is quite different with the model introduced in \cite{ludwig1}\cite{ludwig2}, and as one will see in section 7, our constructions can be easily generalized to the whole $\mathrm{CI}$ class with any possible topological invariants.

\section{Quantum Anomalous Hall Effect with Arbitrary Chern Numbers}

~~~~~Since the BdG Hamiltonian of a superconductor is analogous to the Bloch Hamiltonian of a band insulator
(with the the superconducting gap being replaced by the band gap),
so there is an analogy between 2d chiral $d+id$ topological superconductors
and the 2d quantum anomalous Hall effects.
Hence, we can apply a likewise construction to build models for QAH effects with arbitrary Chern numbers.

Let' s consider a class of two bands models for QAH effects, with the Bloch Hamiltonian given by
\be H(\bp)=
\(\begin{array}{cc}
m+t|w(p_x,p_y)|^2 & \bar{w}(p_x,p_y)\\
w(p_x,p_y) & -(m+t|w(p_x,p_y)|^2)
\end{array}\),\label{md3}\ee
where $t>0$. And the function $w(p_x, p_y)$ is given by $w(p_x, p_y)=w(p_x+ip_y)$,
here $w(p_x+ip_y)$ is a degree $n$ holomorphic function of the complex momentum $p=p_x+ip_y$,
\be w(p)=A\prod^n_i(p-a_i),\label{w}\ee $A$ and $a_i$ are all complex numbers.
One can write $w(p_x, p_y)$ explicitly as
\be w(p_x, p_y)=A\prod^n_i(p_x+ip_y-a_i).\ee

It is easy to find the eigenvalues of (\ref{md3}), with
\be E(\bp)_{\pm}=\pm\sqrt{(t|w|^2+m)^2+|w|^2}.\ee
If $m\neq 0$, the band gap is open on the whole momentum space.
When $m>0$, the system is topologically trivial, since in this case we can continuously deform $A$ to zero without closing the band gap.
When $m<0$, the system will transit to a topologically nontrivial phase, in fact, in this case $H(\bp)$ (\ref{md3}) is homotopically equivalent to
\be
\(\begin{array}{cc}
|w(p_x,p_y)|^2-1 & 2\bar{w}(p_x,p_y)\\
2w(p_x,p_y) & -(|w(p_x,p_y)|^2-1)
\end{array}\),\label{md3sol}\ee
with the holomorphic function is now given by $w(p_x, p_y)=A(p_x+ip_y)^n$.

The wave function $\Psi(p_x, p_y)$ of the occupied band of the Hamiltonian (\ref{md3sol})
can be easily carried out,
\be\Psi(p_x, p_y)=\(\begin{array}{cc}
 1\\
-w(p_x,p_y)
\end{array}\)/\sqrt{1+|w(p_x,p_y)|^2}.\ee
One can easily calculate the Berry phase ${\mathcal{A}}$ and Berry curvature $\mathcal{F}$,
then the first Chern number can be given by the following formula,
\be c_1=\frac{i}{2\pi}\int_{\bp} \frac{dw\wedge d\bar{w}}{(1+|w|^2)^2}.\label{e}\ee
Here we are integrating over the whole momentum space. By substituting $w(p)=Ap^n$, one can easily get
\be c_1=n.\ee

Therefore, to any given positive number $n$,
we have constructed a concrete QAH effect model (\ref{md3}), with the Hall conductance $\sigma_H=-c_1e^2/h=-ne^2/h$\cite{TKNN}.
And it is easily to see that when $n=1$, the holomorphic function $w$ becomes $w(p_x+ip_y)=A(p_x+ip_y)$, and $|w|^2\propto \bp^2$,
then our model (\ref{md3}) becomes the model of \cite{qam}.

The QAH effects with negative Chern numbers $-n$ can also be realized by the same models (\ref{md3}),
but in this case we should replace the holomorphic function $w(p)$ by an anti-holomorphic function $w(\bar{p})$, with
\be w(p_x, p_y)=B\prod^n_i(p_x-ip_y-b_i),\ee
where $B$ and $b_i$ are all complex numbers.

A notable property of the Chern number formula (\ref{e}) is, it is invariant under the transformation
\be w\rightarrow \frac{1}{w}. \ee
This indicates that under the replacement $w\rightarrow 1/w$, our model (\ref{md3}) will become a new model
(called the dual model. This duality is reminiscent of the Abelian duality of two dimensional quantum field \cite{chernd}.) that is describing the same topological phase. This dual model is
\be\frac{1}{|w|^2}\(\begin{array}{cc}
t+m|w(p_x,p_y)|^2 & {w}(p_x,p_y)\\
\bar{w}(p_x,p_y) & -(t+m|w(p_x,p_y)|^2)
\end{array}\).\label{md3d}\ee
Here the factor $\frac{1}{|w|^2}$ of (\ref{md3d}) is irrelevant for the topology of eigenstates, so can be omitted.
As one can see that, in this dual model, the two parameters $t, m$ are switched, hence this new model is topologically nontrivial if $m>0, t<0$.
Thus, combing the previous conditions $t>0, m<0$ for the topological nontrivial phase of (\ref{md3}),
we can conclude that our model (\ref{md3}) is topologically nontrivial if
\be mt<0.\ee
In the case of $n=1$, this result is well known\cite{QAr}.

\section{New Models for $Z_2$ Topological Insulators}

~~~~~~It is well known that the QAH effects are closely related to the $Z_2$ TR invariant topological insulators\cite{qare}.
Thus, our constructions in last section can also be applied to build a class of new models for $Z_2$ TR invariant topological insulators.

Just like the BHZ model for quantum spin Hall effect,
we can build new models for 2d quantum spin Hall effects,
simply by a direct sum of two copies of independent quantum anomalous Hall models,
\be
\(\begin{array}{cccc}
M(p_x,p_y) & \bar{w}(p_x,p_y) &0&0\\
w(p_x,p_y) & -M(p_x,p_y)&0&0\\
0&0&M(-p_x,-p_y)& w(-p_x, -p_y)\\
0&0&\bar{w}(-p_x,-p_y)&-M(-p_x,-p_y)
\end{array}\),\label{qsh}\ee
where $M(p_x,p_y)=m+t|w(p_x,p_y)|^2$,
and the two diagonal nonzero $2\times 2$
submatrixes are acting on the spin up and spin down subspaces respectively.
The TR symmetry is $T=i\sigma_yK$ (where the Pauli matrix $\sigma_y$ acts on the spinor indices), satisfying $T^2=-1$.

If $m<0$, the Hamiltonian (\ref{qsh})will have a quantized spin current, with quantum number $n$.
As it is well known, this quantum number $n$ is not a topological invariant.
But $n\mod 2$ is indeed topologically invariance\cite{kane1}.
Thus, if $n$ is odd, the system will be in a topologically nontrivial phase.
We can explicitly demonstrate this for the situation of degenerate holomorphic function $w(p)=Ap^n$
(the general situations are homotopically equivalent to this case).
In this case, besides the TR invariance, there is also an inversion symmetry $P$,
\be P=\(\begin{array}{cccc}
1 & 0 &0&0\\
0 & -1 &0&0\\
0&0& 1 & 0  \\
0&0&0 &-1
\end{array}\).\ee
Hence one can use the powerful formula of \cite{kane3} to calculate the $Z_2$ topological invariant of our model (\ref{qsh}).
Since there is only one TR invariant point $\bp=0$,
one can easily see that the $Z_2$ topological invariant is nontrivial, only when $m<0$.

There are some subtleties that may need to be clarified. The original formula of \cite{kane3}
is defined over all TR invariant points of the whole
Brillouin zone, which is a closed manifold. Thus to apply this formula to our models,
we need to take points at the infinity into consideration.
This can be achieved by dividing the Hamiltonian (\ref{qsh}) a positive factor $1+C\bp^{2n}$($C>0$)
(since this factor have nothing to do with the eigenstates),
to make our model be defined on the whole momentum space $S^2$, including the infinity.
Now there are two TR invariant points $\bp=0$ and $\bp=\infty$.
And a little thought shows us that at $\bp=\infty$ the eigenvalue
(for the low lying Kramers doublet) of the inversion $P$ depends on $\sign(t)$ (which we have previously assumed to be 1),
while at $\bp=0$ the eigenvalue of $P$ depends on $\sign(m)$. Thus we can finally get the condition for a topologically nontrivial phase,
which is the well known $mt<0$.

We now turn to construct a large class of new models for 3d TR invariant topological insulators.
We will maintain the inversion symmetry in our constructions, since in this situation we can use the powerful formula of \cite{kane3}
to compute the $Z_2$ topological invariant.

We will begin by defining the holomorphic function $w(p)$ as (\ref{w}), but now we set $n=2k+1$,
and take $w(p)=Ap^{2k+1}$. Then, for any given pair of nonnegative integers $(k,l)$, we can define a model with band
Hamiltonian $H_{(k,l)}$
\be
H_{(k,l)}=\(\begin{array}{cccc}
M(\bp) & A_zp^{2l+1}_z &0& \bar{w}(p_x,p_y)\\
A_zp^{2l+1}_z & -M(\bp)& \bar{w}(p_x,p_y) &0\\
0& w(p_x, p_y) & M(\bp) & -A_zp^{2l+1}_z\\
w(p_x,p_y) & 0& -A_zp^{2l+1}_z &-M(\bp)
\end{array}\),\label{3d1}\ee
where, $A_z$ is real, $M(\bp)=m+t|w(p_x,p_y)|^2+t_zp^{2(2l+1)}_z$,
and we will assume $t>0, t_z>0$ for simplicity.
The basis of (\ref{3d1}) is the same as in the quantum spin Hall effects cases,
that is, the diagonal two $2\times 2$ submatrixes of (\ref{3d1})
are acting on the spin up and spin down subspaces respectively.
The formula of \cite{kane3} can be used to show that, when $m<0$,
the $Z_2$ topological invariant of our model $H_{(k,l)}$ is nontrivial.
And we would like to note that the $H_{(0,0)}$ model is just the model
that can be realized in $\mathrm{Bi_2Se_3}$, $\mathrm{Bi_2Te_3}$ and $\mathrm{Sb_2Te_3}$\cite{TI}.

Now we would like to present our constructions in a different basis
$|+ \uparrow\r, |- \downarrow\r, |+ \downarrow\r, |- \uparrow\r$,
where $+, -$ label the parities of the orbits under inversion $P$.
In this basis, our models for 3d topological insulators can then be rewritten as
\be
H_{(k,l)}=\(\begin{array}{cccc}
M(\bp) & \bar{w}(p_x,p_y) &0& A_zp^{2l+1}_z\\
w(p_x,p_y) & -M(\bp)& A_zp^{2l+1}_z&0\\
0& A_zp^{2l+1}_z & M(\bp) & -w(p_x, p_y)\\
A_zp^{2l+1}_z & 0& -\bar{w}(p_x,p_y)&-M(\bp)
\end{array}\).\label{3d2}\ee

In these two constructions(\ref{3d1})(\ref{3d2}), the matrix of the TR symmetry $T$ is the same, given by
\be T=\(\begin{array}{cccc}
0 & 0 &1&0\\
0 & 0 &0&1\\
-1&0& 0 & 0  \\
0&-1&0 &0
\end{array}\)K.\ee

People may try to take the number $n$ to be an even number, but this will invalid the inversion symmetry,
consequently one will have no formula to use to guarantee the nontrivial topology.

\section{Chiral Edge States}

~~~In this section, we will study the chiral edge modes of the models
(for QAH insulators with arbitrary Chern numbers) that we constructed in section(3).
The same analysis can also be applied to the model for $d+id$ chiral topological superconductor
that we studied in section (2).

For simplicity, we will take the model with $c_1=2$ as an example.
The investigations in section (3) tell us
that this kind of model is homotopically equivalent to
\be H(p_x,p_y)=
\(\begin{array}{cc}
m+|\Delta|^2(p^2_x+p^2_y)^2 & 2\bar{\Delta}(p_x-ip_y)^2\\
2\Delta(p_x+ip_y)^2 & -(m+|\Delta|^2(p^2_x+p^2_y)^2)
\end{array}\).\label{hotm1}\ee
Hence we will focus on the specific model with the Hamiltonian given by (\ref{hotm1}),
but our conclusions can be applied to the whole homotopically class.
As we have discussed, when $m>0$ this model (\ref{hotm1}) will be in topologically trivial phase,
while when $m<0$, it will be in a nontrivial phase with $c_1=2$.

In what follows, we will demonstrate the existence of chiral edge modes when $m<0$, by solving the model in half space with an open boundary condition.
As it is well known, see \cite{witten1}\cite{witten2} for examples, these gapless chiral edge modes must be exist for consistency.
The reason is that, in the $m<0$ phase, the bulk low energy effective theory of (\ref{hotm1}) can be described by a Chern-Simons theory at level $k=c_1=2$\cite{pt1}, this theory is anomaly at the boundary, this anomaly should be cancelled by the chiral anomaly of some chiral edge modes. This bulk boundary correspondence tells us that the numbers of chiral edge modes in our model (\ref{hotm1}) should be $2$ (when $m<0$).
We will show that this is exactly the case, but the detail mechanism is quite different with the $c_1=1$ case (see \cite{QAr}).

Assuming that we are solving the model (\ref{hotm1}) in half space with $x\geq 0$. Hence $p_y$ remains a good quantum number,
but the $p_x$ in (\ref{hotm1}) should be replaced by operator $\hat{p}_x=-i\p/\p x$.
Now, let's firstly consider the case of $p_y=0$. From (\ref{hotm1}) we can see that in this case
the corresponding Hamilton operator $\hat{H}_0$ should be
\be \hat{H}_0=\(\begin{array}{cc}
m+|\Delta|^2{\p^4}/{\p x^4} & -2\bar{\Delta}{\p^2}/{\p x^2}\\
-2\Delta{\p^2}/{\p x^2} & -(m+|\Delta|^2{\p^4}/{\p x^4})
\end{array}\).\ee
We will search for the normalizable zero mode (zero energy) solution $\psi(x)$ which satisfies
\be\hat{H}_0\psi(x)=0.\label{zero}\ee
The requirement of normalizability will impose a boundary condition
\be\psi(\infty)=0.\label{bc1}\ee
Further more, $\hat{p}_x$ should be a Hermitian operator, this will impose another boundary condition
\be\psi(0)=0.\label{bc2}\ee

Assuming that the zero mode $\psi(x)$ can be written as a linear superposition of
the functions of the form $e^{\lambda x}\phi$,
where $\phi$ is a two components constant vector.
By using the zero mode equation (\ref{zero}) we can have
\be
\(\begin{array}{cc}
m+|\Delta|^2\lambda^4 & -2\bar{\Delta}\lambda^2\\
-2{\Delta}\lambda^2 & -(m+|\Delta|^2\lambda^4)
\end{array}\)\phi=0.\label{100}\ee
Now we denote the complex parameter as $\Delta=|\Delta|e^{i\delta}$, then we can rewrite (\ref{100}) as
\be\((m+|\Delta|^2\lambda^4)-2i\lambda^2|\Delta|
\tau_{\delta}\)\phi=0,\label{phieq}\ee
where $\tau_{\delta}$ is a $2\times 2$ Hermitian matrix $\tau_{\delta}=\(\begin{array}{cc}
0 & -ie^{-i\delta}\\
ie^{i\delta} & 0
\end{array}\)$.
We will denote the eigenvalues of $\tau_{\delta}$ as $s$, obviously $s=\pm 1$.

As one can easily see that, equation (\ref{phieq}) means $\phi$ must be one of the eigenstates of
$\tau_{\delta}$, denoted as $\phi_{(s)}$, with eigenvalue $s$,
\be\tau_{\delta}\phi_{(s)}=s\phi_{(s)}.\label{tau}\ee
Hence, equation (\ref{phieq}) becomes
\be|\Delta|^2\lambda^4-2i|\Delta|s\lambda^2+m=0.\label{ll}\ee
The roots are
\be(\lambda^2)_{\pm}=\frac{s\pm\sqrt{1+m}}{|\Delta|}i.\label{l1}\ee
It is easy (from (\ref{ll})) to see that, the roots for the case of $s=-1$ are just the complex conjugations of the roots for $s=1$.
For any one of the two values of $s$, we have four roots for $\lambda$, two of these four roots have positive real parts, and the other two have negative real parts. Only the roots with negative real parts can satisfy the boundary condition (\ref{bc1}), we will denote these two roots as $\lambda_1, \lambda_2$. To satisfy the boundary condition (\ref{bc2}) at the same time, the zero mode $\psi(x)$ must take the form of $\psi_s(x)$, with
\be\psi_s(x)=(e^{\lambda_1x}-e^{\lambda_2x})\phi_{(s)}.\ee

Now let's consider the situation of $p_y\neq 0$. We will take $p_y$ as a small perturbation parameter.
Obviously, to satisfy the zero order (of $p_y$) equation (\ref{zero}), the wave functions of edge modes must be
\be\psi_{p_y,s}(x,y)=\psi_s(x)e^{ip_yy}.\ee
Moreover, from the full Hamiltonian (\ref{hotm1}),
we can read off the first order perturbation $H'$
\be H'=4\(\begin{array}{cc}
0 & -i\bar{\Delta}\\
i\Delta & 0
\end{array}\)p_y\hat{p_x}.\ee
In what follows we will calculate the first order correction $\l\psi_{p_y,s}|H'|\psi_{p_y,s}\r$ to the energy of the edge modes.

Firstly, we notice
\be\Delta_s={\phi_{(s)}^{\dag}}\(\begin{array}{cc}
0 & -i\bar{\Delta}\\
i\Delta & 0
\end{array}\)\phi_{(s)}=|\Delta|{\phi_{(s)}^{\dag}}\tau_{\delta}\phi_{(s)}=s|\Delta|.\ee

Then, in the situation of $m>0$, simple analysis to (\ref{l1}) tells us, what ever $s=1$ or $s=-1$,
the two relevant roots $\lambda_1, \lambda_2$ are always proportional to
\be\lambda_1\varpropto -\cos(\pi/4)+i\sin(\pi/4), \lambda_2\varpropto -\cos(\pi/4)-i\sin(\pi/4).\ee
A few calculations show us, in this situation, $\l\psi_{s}|\hat{p_x}|\psi_{s}\r=0$.
Hence, when $m>0$, there are gapless excitations localized at the boundary,
but in this situation the dispersion relation of these gapless edge modes must be in the form of $\varepsilon(p^2_y)$
(since the first order correction is zero).
Thus they can not be chiral modes, just as required by the bulk boundary correspondence.

In the situation of $m<0$, detail analysis to (\ref{l1}) shows us:
for $s=1$, both the two relevant roots $\lambda_1, \lambda_2$ are at
the lower half of the complex plane, thus have the forms of
\be\lambda_1=-a-bi, \lambda_2=-c-di,\label{l1l2}\ee
where $a,b,c,d$ are all positive numbers. In this case, detail calculations show us $\l\psi_{s}|\hat{p_x}|\psi_{s}\r$ is a negative number,
proportional to $-(bc+ad)$. Noticing that $\Delta_{s=1}=|\Delta|$, hence $\l\psi_{p_y,s}|H'|\psi_{p_y,s}\r$ is proportional to $-p_y$.
Thus, in the case of $s=1$, there is a left moving chiral edge mode, with the dispersion relation
\be\varepsilon=-v_Fp_y.\ee
 For the case of $s=-1$, detail analysis shows us, both the two relevant roots $\lambda_1, \lambda_2$ are at the upper
 half of the complex plane. These two roots are the complex conjugations of (\ref{l1l2}), thus have the forms of
\be\lambda_1=-a+bi, \lambda_2=-c+di.\ee
Detail calculation shows us $\l\psi_{s}|\hat{p_x}|\psi_{s}\r$ is now positive,
in fact it is proportional to $(bc+ad)$. Noticing that $\Delta_{s=-1}=-|\Delta|$,
hence we also have $\l\psi_{p_y,s}|H'|\psi_{p_y,s}\r\propto -p_y$. We thus have another left moving chiral edge mode,
also with the dispersion relation
\be\varepsilon= -v_Fp_y.\ee
In summary, when $m<0$, we have two chiral edge modes,
just as required by the bulk boundary correspondence.
\\

If what we are dealing with is the model for chiral $d+id$ topological superconductor that we constructed in section 2,
since the BdG Hamiltonian for this case has the same form as (\ref{hotm1}),
thus the analysis for the edge modes will be identical with our above analysis
(The corresponding equation is the BdG equation with the BdG Hamiltonian acting on the Nambu spinor $\Psi(x,y)=(c_{\uparrow}(x,y),c^{\dagger}_{\downarrow}(x,y))^{T}$ in coordinate space ). But in the present situation,
we have a particle hole symmetry $C$, which is anticommuting with the matrix $\tau_{\delta}$.
Thus, we can choose the two solutions $\phi_{(s)}$ of (\ref{tau}) to satisfy
\be\phi_{(-1)}=C\phi_{(+1)}, \phi_{(+1)}=-C\phi_{(-1)}.\ee
Furthermore, as we have discussed,
the four roots (for $\lambda$) of the $s=1$ case and that of the $s=-1$ case are the complex conjugations of each other.
Thus we can choose the two zero modes $\psi_s(x)$ to satisfy
\be \psi_{-1}(x)=C\psi_{+1}(x), \psi_{+1}(x)=-C\psi_{-1}(x).\ee
Then, we can define the operators of edge modes as
\be\alpha^{\dagger}_{p_y, \uparrow}&=&\int_{x>0} dxdy\Psi^{\dagger}(x,y)\cdot\psi_{+1}(x)e^{ip_yy}\\
\alpha_{-p_y, \downarrow}&=&\int_{x>0} dxdy\Psi^{\dagger}(x,y)\cdot\psi_{-1}(x)e^{-ip_yy}.\ee
These two operators can create two spin up edge fermions, and since charge conjugation $C$ exchanges the two,
hence these two fermions are the "anti-particles" of each other.

As we have shown that, in the topologically nontrivial phase, the gapless edge modes are chiral, hence the corresponding second quantized Hamiltonian for these edge modes can be written as
\be H_{sf}= -v_F\int p_y\(\alpha^{\dagger}_{p_y, \uparrow}\alpha_{p_y, \uparrow}+\alpha^{\dagger}_{p_y, \downarrow}\alpha_{p_y, \downarrow}\).\ee
The chiral nature of these edge modes indicate that they are immune from backscattering,
and can not be gapped by small perturbations or disorders.
In fact these chiral edge modes are protected by the particle hole symmetry
and the nontrivial bulk topology $c_1=2$.

Since $\alpha_{p_y, \downarrow}$ and $\alpha_{p_y, \uparrow}$ are charge conjugated to each other, they can form a single complex chiral edge fermion field.
Hence, the number of chiral edge modes of a $c_1=2$ $d+id$ topological superconductor is equal to the number of edge modes of
a $c_1=1$ quantum Hall effect. The reason is that, in the situations of topological superconductors,
the positive energy states and the negative energy states of BdG Hamiltonian are describing the same physical degrees of freedom,
thus the degrees of freedom of a topological superconductor are in general half of the degrees of freedom
of the quantum Hall effect with the same topological invariant.

Just as in the cases of QAH effects, where the Chern numbers are related to a physical observable (the Hall conductance, by the TKNN formula),
here, for the cases of 2d class $\mathrm{C}$ topological superconductors,
the Chern numbers are also related to a physical observable, the Hall spin conductance.
Thus the topological classification to 2d class $\mathrm{C}$ topological superconductors
may be extended to the interaction situations. This is quite different
with the situations of conventional 3d TR invariant topological superconductors, where
interactions can change the $\ZZ$ classification of the free fermion theories \cite{161}\cite{162}\cite{witten2}.

\section{Surface States of New Models for $Z_2$ Topological Insulators}

~~~~In this section we will study the surface states of the models $H_{(k,l)}$ that we constructed in section 4. For simplicity,
we will take the case of $H_{(k,0)}$ as an example, and then generalize the conclusions to the general $H_{(k,l)}$ models.

We will solve the model $H_{(k,0)}$ at half space with coordinate $z\geq 0$.
We firstly study the zero energy normalizable solutions $\psi(z)$,
with $p_x=p_y=0$. The surface states will then be constructed as $\psi_{p_x, p_y}(x,y,z)=\psi(z)e^{i(p_xx+p_yy)}$.
Finally, we will compute the effective Hamiltonian of these surface states.

To the zero and first orders of $p_x, p_y$,
the Hamiltonian $H_{(k,0)}$ can be divided into two parts, $H_{(k,0)}=\hat{H}_0+H'$,
the zero order part $\hat{H}_0$ is given by (with the replacement $p_z\rightarrow -i\p/\p z$)
\be \hat{H}_0=\(\begin{array}{cccc}
M(\p/\p z) & 0 &0& -iA_z\p/\p z\\
0 & -M(\p/\p z) & -iA_z\p/\p z &0\\
0& -iA_z\p/\p z & M(\p/\p z) & 0\\
-iA_z\p/\p z & 0& 0& -M(\p/\p z)
\end{array}\),\ee
where $M(\p/\p z)=m-t_z\p^2/\p z^2$. And the first order part $H'$ is given by
\be H'=\(\begin{array}{cccc}
0 & \bar{w}(p_x,p_y) &0& 0\\
w(p_x,p_y) & 0 & 0&0\\
0& 0 & 0 & -w(p_x, p_y)\\
0 & 0& -\bar{w}(p_x,p_y)&0
\end{array}\).\ee

The zero mode should satisfy $\hat{H}_0\psi(z)=0$, with the open boundary conditions
\be\psi(0)=0,\  \psi(\infty)=0.\label{opencondition}\ee
Assuming that $\psi(z)$ is a linear superposition of the wave functions of the form $e^{\lambda z}\phi_s$,
where $\phi_s$ satisfies the following boundary condition,
\be i\(\begin{array}{cccc}
0 & 0 &0& 1\\
0 & 0& -1 &0\\
0& 1 & 0 & 0\\
-1 & 0& 0 &0
\end{array}\)\phi_s=s\phi_s,\label{phis}\ee
with $s=\pm 1$, and $\lambda$ satisfies the equation
\be t_z\lambda^2+sA_z\lambda-m=0.\label{lr1}\ee
These two equitations (\ref{phis})(\ref{lr1}) are derived from the zero mode equation $\hat{H}_0\psi(z)=0$.
Further more, just as in the discussions for 2d chiral edge modes, of all the roots of (\ref{lr1}), only that roots
with negative real parts can lead normalizable solutions.

Since $T^2=-1$, hence if the zero energy solutions $\psi(z)$ do exist, they will form Kramers pairs.
Noticing that $T$ is commuting with the eigenstate equation (\ref{phis}),
thus every single zero mode Kramers pair satisfies a single boundary condition with a single fixed value $s$.

Detail analysis shows us, when $m>0$, the zero mode solution does not exist.
But when $m<0$, there is a single zero mode Kramers pair. When $A_z>0$,
this Kramers pair should satisfy the boundary condition of $s=1$,
otherwise when $A_z<0$, it should satisfy the boundary condition of $s=-1$.

But if what we are considering are the general models $H_{(k,l)}$, with $l\geq 1$, there will be some additional subtleties.
In these situations, the zero mode solutions $\psi(z)$ can exist even when $m>0$. But in this case, whatever the boundary conditions are, $s=1$ or $s=-1$,
the numbers of independent zero mode Kramers pairs must be even.
But when $m<0$, the numbers of independent zero mode Kramers pairs are always odd,
and in generally take different values for the two different boundary conditions, $s=1$ or $s=-1$.

Now we turn to the situation of $p_x, p_y\neq 0$.
Obviously, each zero mode Kramers pair now gives a gapless two components surface excitation. If one choose the basis as
$(|\uparrow\r, |\downarrow\r)$, one can write down the effective Hamiltonian for this surface excitation
\be H_{surf}(p_x,p_y)=
\(\begin{array}{cc}
0 & i\bar{w}(p_x,p_y)\\
-i{w}(p_x, p_y) & 0
\end{array}\).\label{sf}\ee
Due to the TR invariance, the diagonal terms of this effective Hamiltonian (\ref{sf}) are zero.

Remembering that $w(p)=Ap^{2k+1}$, hence for the cases of $k\geq 1$, what (\ref{sf}) is describing is not a single Dirac fermion,
but a degenerate Dirac cone.
If fact, one can continuously deform $w(p)$, maintaining TR symmetry, to a
nondegenerate form $w(p)=Ap\prod^k_i(p-a_i)(p+a_i)$. The boundary effective Hamiltonian (\ref{sf})
with this nondegenerate $w(p)$ will have $n=2k+1$ Dirac points in the momentum space of the surface.
These $n=2k+1$ Dirac points are describing $n=2k+1$ massless Dirac fermions.
Thus, each Kramers pair of the zero modes gives out
$n=2k+1$ massless Dirac fermions at the surface.

Hence, the even Kramers pairs in the case of $m>0$ must give out even numbers of surface Dirac fermions,
while the odd numbers of Kramers pairs in the case of $m<0$ must give out odd numbers of surface Dirac fermions.
But, if the numbers of surface Dirac fermions are even, one can always add TR invariant mass terms pairwise,
such as $im\(\psi^{\dagger}_1\sigma_z\psi_2-\psi^{\dagger}_2\sigma_z\psi_1\)$
($\psi_1$, $\psi_2$ are two of these surface fermions).
Consequently, the surface excitations can be totally gapped. Thus, the phase of $m>0$ must be topologically trivial.
On the other hand, when the numbers of surface Dirac fermions are odd, there are always odd numbers of
gapless surface fermions surviving any mass deformations, thus, the phase of $m<0$
have a nontrivial $Z_2$ topology.

If what we are considering is a thin film of 3d topological insulator, in this situation, we will have two surfaces, the top surface and the bottom surface. The surface states at each surface can now tunnel to the opposite surface. The effective Hamiltonian of these two coupled surface states can be written in the bonding and antibonding basis $|+ \uparrow\r, |- \downarrow\r, |+ \downarrow\r, |- \uparrow\r$\cite{qa3d}, as
\be\(\begin{array}{cccc}
M(p_x,p_y) & i\bar{w}(p_x,p_y)&0&0\\
-i{w}(p_x, p_y) & -M(p_x,p_y) &0&0\\
0&0&M(p_x,p_y) & -i{w}(p_x,p_y)\\
0&0&i\bar{w}(p_x,p_y)&-M(p_x,p_y)
\end{array}\),\label{transition}\ee
where $M(p_x,p_y)=m+t|w(p_x,p_y)|^2$.
This Hamiltonian has a roughly physical interpretation. Firstly, by the effective Hamiltonian (\ref{sf}), we can write down the tunneling amplitude
\be{i\bar{w}}: |- \downarrow\r\rightarrow |+ \uparrow\r,\ \ -i{w}: |+ \uparrow\r\rightarrow|- \downarrow\r\\
{-iw}: |- \uparrow\r\rightarrow|+ \downarrow\r,\ \ {i\bar{w}}: |+ \downarrow\r\rightarrow|- \uparrow\r,\ee
where the tunnelings can only happen between the states with opposite parities,
since the amplitudes $w,\bar{w}$ are odd parities.
These amplitudes of tunnelings give the off diagonal terms of (\ref{transition}).
And the term proportional to $|w|^2$ in the diagonal terms may be understood as the amplitude of two successive tunnelings, forth and back.
As proposed by \cite{qa3d}, this system of thin film (of 3d topological insulator) may be used to realized the 2d quantum anomalous Hall effect with any
$n=2k+1$ Chern numbers.

\section{Surface States of $d$ Wave Topological Superconductors}

~~~~~In this section we will investigate the surface states of 3d $d$ wave topological superconductor.
We will consider a general class of models for 3d class $\mathrm{CI}$ topological superconductors (classified by $2\ZZ$)
with topological invariants $n=2k$.
The model for 3d $d$ wave topological superconductor that we constructed in section 2 is
just the special case with $n=2$.

The BdG Hamiltonians of this general class of models can be given as
\be
H^{(k,l)}_{BdG}=\(\begin{array}{cccc}
E(\bp) & \bar{w}(p_x,p_y) &0& -\Delta_z p^{2l+1}_z\\
w(p_x,p_y) & -E(\bp) & \Delta_z p^{2l+1}_z&0\\
0& \Delta_z p^{2l+1}_z & E(\bp) & w(p_x, p_y)\\
-\Delta_z p^{2l+1}_z & 0& \bar{w}(p_x,p_y)&-E(\bp)
\end{array}\),\ee
where $w(p_x,p_y)=\Delta p^{n}=\Delta p^{2k}$, $p=p_x+ip_y$ is the complex momentum, and $E(\bp)=t|w(p_x, p_y)|^2+t_zp^{2(2l+1)}_z-\mu $.

Just as what we have done in last section, we will solve these models at the half space with $z\geq 0$.
We will firstly study the normalizable zero energy wave $\psi(z)$(with $p_x=p_y=0$).
$\psi(z)$ should satisfy the zero mode equation
$H^{(k,l)}_{BdG}(p_x=p_y=0, p_z=-i\p/\p z)\psi(z)=0$
and the open boundary conditions (\ref{opencondition}).

Assuming that $\psi(z)$ is a linear superposition of the functions of the form $e^{\lambda z}\phi_s$,
here $\phi_s$ satisfies the following boundary condition
\be i\(\begin{array}{cccc}
0 & 0 &0& -1\\
0 & 0& -1 &0\\
0 & 1 & 0 & 0\\
1 & 0& 0 &0
\end{array}\)\phi_s=s\phi_s,\ee
and $\lambda$ satisfies,
\be t_z(\lambda^{2l+1})^2+(-)^ls\Delta_z(\lambda^{2l+1})+\mu=0.\ee
These equations can be easily derived from the zero mode equation.

Since $C^2=-1$, hence there is a 'Kramers degeneracy'. Thus the zero modes $\psi(z)$, if exist,
must form Kramers pairs. The situation is much like what we have studied in last section.
Detail analysis shows us, when $l=0$, there is a zero mode pair only when $\mu>0$. If $\Delta_z>0$,
this zero mode pair will satisfy the boundary condition of $s=1$, while if $\Delta_z<0$, this pair will satisfy the boundary condition of $s=-1$.
But when $l\geq 1$, there are always zero modes, even if $\mu<0$.
When $\mu<0$, the numbers of these zero mode Kramers pairs are always even,
no matter $s=1$ or $s=-1$.
While when $\mu>0$, the numbers of the zero mode Kramers pairs are always odd, no matter $s=1$ or $s=-1$
(the two choices have different numbers of zero mode pairs).

Having found the zero modes $\psi(z)$,
we can now construct the surface states $\psi_{p}$ by using wave function $\psi_{p_x, p_y}(x,y,z)=\psi(z)e^{i(p_xx+p_yy)}$,
with \be\psi^{\dagger}_{p}=\int dxdydz\Psi^{\dagger}(x,y,z)\psi_{p_x, p_y}(x,y,z),\ee
where $\Psi(x,y,z)$ is the four components Nambu spinor introduced in section 2.
Each zero mode Kramers pair will give a two components surface excitation $\psi_{p}=(\gamma_{p\uparrow}, \gamma_{p\downarrow})^{T}$, with
\be\gamma_{p\uparrow}=c_{p\uparrow}+c^{\dagger}_{-p\uparrow},\
\gamma_{p\downarrow}=c_{p\downarrow}+c^{\dagger}_{-p\downarrow},\ee
where $c_{p\uparrow}, c_{p\downarrow}$ are labeled by the surface momentum.
Obviously, $\psi_{p}$ satisfy the Majorana condition
\be\psi_{-p}=\psi^{\ast}_{p}=(\gamma^{\dagger}_{p\uparrow}, \gamma^{\dagger}_{p\downarrow})^{T}.\ee
And the particle hole symmetry acts as \be C\gamma_{p\uparrow}C^{-1}=\gamma_{p\downarrow}, C\gamma_{p\downarrow}C^{-1}=-\gamma_{p\uparrow}.\ee
Since these two fermions have different charge conjugation parities, they are not Majorana fermions.

To further investigate the topological natures of our models by using the bulk boundary correspondence,
we need to study the actions of $T$ at the surface modes, which is
\be T\psi_p=\pm\(\begin{array}{cc}
0 & 1\\
1 & 0
\end{array}\)\psi_{-p}.\label{T}
\ee
Where the $\pm$ signatures in (\ref{T}) mean that
for each surface doublet $\psi_{i}$ (corresponding to the $i^{th}$ zero mode Kramers pair),
there are two different choices (distinguished by the $\pm$ signatures) for the $T$ action. Thus, different surface doublets,
such as $\psi_{i}, \psi_{j}, i\neq j$, can have different $T$ actions (with different signs).
Consequently, although the mass term $m_i\psi^{\dagger}_i\sigma_z\psi_i$ is TR broken,
but for two different surface doublets $\psi_{i}, \psi_{j}$ with different $T$ actions,
we can form a TR invariant mass term
$m_{ij}(\psi^{\dagger}_i\sigma_z\psi_j
+\psi^{\dagger}_j\sigma_z\psi_i)$.

Thus the even numbers of zero mode Kramers pairs, as in the case of $\mu<0$, are in fact topologically equivalent to 0 Kramers pair,
since we can impose $T$ actions pairwise,
to let all the corresponding surface excitations be totally gapped
by some appropriate TR invariant mass deformations.
While the odd numbers of Kramers pairs, as in the case of $\mu>0$,
are topologically equivalent to a single Kramers pair, since
there are always odd numbers of surface doublets surviving any permissible mass deformations.
The second quantized effective Hamiltonian for this surviving surface doublet is
\be H_{sf}=\int \(w(p)\gamma_{-p\downarrow}\gamma_{p\uparrow}+\bar{w}\gamma_{-p\uparrow}\gamma_{p\downarrow}\).\label{dwavesf}\ee

Since $w(p)=\Delta p^{n}$ with $n=2k>1$, (\ref{dwavesf}) does not describe a single Dirac fermion,
what it is describing is $2k$ massless Dirac fermions. To see this clearly, we can continuously deform $w(p)$,
maintaining TR symmetry and particle hole symmetry, to a factorized form $w(p)=\Delta\prod^k_i(p-a_i)(p+a_i)$, which gives $2k$ Dirac points at the surface momentum space.
Since these $2k$ Dirac fermions come from the same zero mode Kramers pair, they can not have different $T$ actions.
A little thought shows us, in this situation any mass terms will either invalid TR symmetry, or invalid particle hole symmetry.
So we end up with $n=2k$ gapless surface excitations. This means $n=2k$ is the topological invariant of our system.

Now we can conclude that, the $\mu<0$ phases of our models are always topological trivial, while the phases of $\mu>0$
are topologically nontrivial and have topological invariants $n=2k$, just as required by the $2\ZZ$ classification
of 3d class $\mathrm{CI}$ topological superconductors.
When $k=1$, we will get the case of 3d $d$ wave topological superconductor.

\section{Conclusion and Discussions}

~~~~~In this paper, we have constructed nontrivial models for the 2d class $\mathrm{C}$ topological superconductors, and for the 3d class $\mathrm{CI}$ topological superconductors. Our models can realize these two class of topological superconductors with any $2\ZZ$ topological invariants. In the simplest cases, our models describe the $d+id$ topological superconductor (in the case of 2d), and the $d$ wave topological superconductor (in the case of 3d). We also constructed models for quantum anomalous Hall effects with arbitrary large Chern numbers. Besides, we give a class of new models for the $Z_2$ topological insulators, and in the simplest cases, our models turn out to be the well known models for $Z_2$ topological insulators.

We have studied the chiral edge states or gapless surface states for the topologically nontrivial phases of our models.
And we found some novel mechanisms that make the numbers of boundary states are always in agreement with the nontrivial bulk topology, just as required by the bulk boundary correspondence. This provides nontrivial evidences for the self consistencies of our theories.

The studying for the chiral edge modes of quantum anomalous Hall effect with arbitrary large Chern numbers
may have potential applications in future electronic devices,
because the chiral edge modes can transport electric current without dissipations,
and the only resistances are the contact resistances,
these contact resistances can be effectively reduced by the edge modes with large Chern numbers.

Another mechanism to realize dissipationless transport at room temperature is, of course,
using high $T_c$ superconductors. Thus it is natural to try to
combine the chiral characteristic of the edge modes of quantum Hall effects with the high $T_c$ superconductors.
In fact this is the main motivation that prompts
us to investigate the chiral $d+id$ topological superconductor.
Moreover, as we have seen, the chiral edge modes of chiral $d+id$ topological superconductor
can carry quantized spin current, this may enable potential applications in spintronics.

\section{Acknowledgements}

The authors acknowledge the support of the Doctoral Startup Package Fund of East China University of Technology (No. DHBK201203).


\begin{thebibliography}{99}
\bibitem{TI1}
M. Z. Hasan and C. L. Kane, Rev. Mod. Phys. 82, 3045 (2010).
\bibitem{TI2}
X.-L. Qi and S.-C. Zhang, Rev. Mod. Phys. 83, 1057 (2011).
\bibitem{TI3}
M. Z. Hasan and J. E. Moore, Ann. Review. Condensed Matter Physics 2, 55-78 (2011)
\bibitem{kane1}
C. L. Kane, and E. J. Mele., Physical Review Letters 95.14 (2005): 146802
\bibitem{kane2}
L. Fu, and C. L. Kane., Physical Review B 74.19 (2006): 195312
\bibitem{kane3}
L. Fu, and C. L. Kane., Physical Review B 76.4 (2007): 045302
\bibitem{3d1}
L. Fu, C. L. Kane, and E. J. Mele., Physical Review Letters 98.10 (2007): 106803
\bibitem{3d2}
J. E. Moore, and L. Balents., Physical Review B 75.12 (2007): 121306
\bibitem{pt1}
X.-L. Qi, T. L. Hughes, and S.-C. Zhang, Phys. Rev. B 78, 195424 (2008).
\bibitem{pt21}
A. Kitaev, Periodic Table For Topological Insulators And Superconductors, arXiv:0901.2686.
\bibitem{pt2}
A. P. Schnyder, S. Ryu, A. Furusaki, and A. W. W. Ludwig, Physical Review B, 78.19 (2008), 195125.
\bibitem{pt3}
S. Ryu, A. P. Schnyder, A. Furusaki, and A. W. W. Ludwig, New Journal of Physics 12.6 (2010): 065010.
\bibitem{read}
N. Read and D. Green, Physical Review B 61.15 (2000): 10267.
\bibitem{az1}
M. R. Zirnbauer, J. Math. Phys. 37, 4986 (1996).
\bibitem{az2}
A. Altland and M. R. Zirnbauer, Phys. Rev. B 55, 1142 (1997).
\bibitem{dwave1}
R. B. Laughlin, Physical Review Letters 80.23 (1998): 5188.
\bibitem{dwave11}
T. Senthil, J. B. Marston, and M. P. Fisher, Physical Review B 60.6 (1999): 4245.
\bibitem{dwave2}
B. Horovitz, and A. Golub, Physical Review B 68.21 (2003): 214503.
\bibitem{dwave20}
M. L. Kiesel, C. Platt, W. Hanke, D. A. Abanin, and R. Thomale, Physical Review B 86.2 (2012): 020507.
\bibitem{dwave21}
M. L. Kiesel, C. Platt, W. Hanke, and R. Thomale, Physical review letters 111.9 (2013): 097001.
\bibitem{dwave3}
M. H. Fischer, T. Neupert, C. Platt, A. P. Schnyder,..., and M. Sigrist, Physical Review B 89.2 (2014): 020509.
\bibitem{dwave4}
R. Nandkishore, L. S. Levitov, and A. V. Chubukov, Nature Physics 8.2 (2012): 158-163.
\bibitem{QAr}
C. X. Liu, S. -C. Zhang, and X. -L. Qi, The quantum anomalous Hall effect, arXiv:1508.07106 (2015).
\bibitem{haldane}
F. D. M. Haldane, Physical Review Letters 61.18 (1988): 2015.
\bibitem{qa}
C.-Z. Chang, J. Zhang, X. Feng, J. Shen, Z. Zhang,
M. Guo, K. Li, Y. Ou, P. Wei, L.-L. Wang, Z.-Q. Ji,
Y. Feng, S. Ji, X. Chen, J. Jia, X. Dai, Z. Fang, S.-C.
Zhang, K. He, Y. Wang, L. Lu, X.-C. Ma, and Q.-K.
Xue, Science 340, 167 (2013).
\bibitem{hc0}
H. Jiang, Z. Qiao, H. Liu, and Q. Niu, Physical Review B 85.4 (2012): 045445.
\bibitem{hc1}
W. Jing, B. Lian, H. Zhang, Y. Xu, S. -C. Zhang, Physical Review letters 111.13 (2013): 136801
\bibitem{hc2}
C. Fang, M. J. Gilbert, and B. A. Bernevig, Physical Review Letters 112.4 (2014): 046801
\bibitem{hc3}
S. A. Skirlo, L. Lu, and M. Soljacic, Physical Review Letters 113.11 (2014): 113904
\bibitem{hc4}
S. A. Skirlo, L. Lu, Y. Igarashi, Q. Yan, J. Joannopoulos and M. Soljacic, Physical Review letters 115.25 (2015): 253901
\bibitem{qam}
X.-L. Qi, Y.-S. Wu, and S.-C. Zhang, Physical Review B 74.8 (2006): 085308.
\bibitem{qa2d}
C.-X. Liu, X.-L. Qi, X. Dai, Z. Fang, and S.-C. Zhang, Physical review letters 101.14 (2008): 146802.
\bibitem{qa3d}
R. Yu, W. Zhang, H. J. Zhang, S. C. Zhang, X. Dai, and Z. Fang, Science 329.5987 (2010): 61-64.
\bibitem{qare}
J. Wang, B. Lian, S. C. Zhang, Physica Scripta 2015.T164 (2015): 014003.
\bibitem{BHZ}
B. A. Bernevig, T. L. Hughes, and S.-C. Zhang, Science 314.5806 (2006): 1757-1761.
\bibitem{TI}
H. Zhang, C. X. Liu, X. L. Qi, Z. Fang, S. C. Zhang, Nature physics 5.6 (2009): 438-442.
\bibitem{vivo}
G. E. Volovik, Journal of Experimental and Theoretical Physics Letters 66.7 (1997): 522-527.
\bibitem{TKNN}
D. J. Thouless, M. Kohmoto, M. P. Nightingale, and
M. den Nijs, Physical Review Letters 49.6 (1982): 405.
\bibitem{ludwig1}
A. P. Schnyder, S. Ryu, A. W. W. Ludwig, Physical review letters 102.19 (2009): 196804.
\bibitem{ludwig2}
A. P. Schnyder, P. M. R. Brydon, D. Manske, C. Timm, Physical Review B 82.18 (2010): 184508.
\bibitem{chernd}
T. Chern, Vortex Operator and BKT Transition in Abelian Duality, Physica C (2016), doi:10.1016/j.physc.2016.01.006
\bibitem{witten1}
E. Witten, Three Lectures On Topological Phases Of Matter, arXiv:1510.07698 (2015).
\bibitem{witten2}
E. Witten, Fermion Path Integrals And Topological Phases, arXiv:1508.04715 (2015).
\bibitem{161}
M. A. Metlitski, L. Fidkowski, X. Chen, and A. Vishwanath, Interaction Effects On 3d
Topological Superconductors: Surface Topological Order From Vortex Condensation,
the 16-Fold Way, And Fermionic Kramers Doublets, arXiv:1406.3032.
\bibitem{162}
A. Kitaev, Homotopy-Theoretic Approach to SPT Phases in Action: $\ZZ_{16}$ Classification
of Three-Dimensional Superconductors, lecture notes available at http://www.ipam.ucla.edu

\end{thebibliography}
\end{document}